\documentclass[twocolumn,amsmath,amssymb,aps,prl]{revtex4-1}

\usepackage{graphicx}% Include figure files
\usepackage{dcolumn}% Align table columns on decimal point
\usepackage{bm}% bold math
\usepackage{xcolor}
\newcommand{\be}{\begin{equation}}
\newcommand{\ee}{\end{equation}}
\newcommand{\bea}{\begin{eqnarray}}
\newcommand{\eea}{\end{eqnarray}}

\usepackage{mathtools}
\begin{document}

\title{Quantum response theory for nonequilibrium steady states}

\author{Michael Konopik}
\author{Eric Lutz}
%\affiliation{Department of Physics, Friedrich-Alexander-Universit\"at Erlangen-N\"urnberg, D-91058 Erlangen, Germany}
\affiliation{Institute for Theoretical Physics I, University of Stuttgart, D-70550 Stuttgart, Germany}

%\affiliation{Department of Physics, Friedrich-Alexander-Universit\"at Erlangen-N\"urnberg, D-91058 Erlangen, Germany}
%\affiliation{Institute for Theoretical Physics I, University of Stuttgart, D-70550 Stuttgart, Germany}
%

\begin{abstract}
We develop a general framework for the  steady-state response of dissipative quantum systems. We concretely derive three different, but equivalent, forms of the quantum response function. We  discuss for each of them the role of the noncommutativity of quantum operators and establish links to the Kubo response theory for closed quantum systems. We show in particular that the equilibrium response  vanishes for perturbations that commute with the unperturbed Hamiltonian while the steady-state response does not, highlighting the profound difference between the two.
 \end{abstract}

\maketitle

Response theory is a cornerstone of statistical physics. For equilibrium systems, the fluctuation-dissipation theorem connects the response to a weak external perturbation to the unperturbed correlation function between spontaneous fluctuations \cite{nyq28,cal51,Kub57}.  It offers a powerful tool to analyze general transport properties in numerous areas, from  hydrodynamics to {many-body} and condensed-matter physics \cite{Kub66,han82,Zwa01,mar08}. The fluctuation-dissipation relation has been derived for classical and  closed quantum systems \cite{Kub66,han82,Zwa01,mar08}. It is known to break down for nonequilibrium systems when detailed balance is not obeyed.

In the past decade, the fluctuation-dissipation theorem has been successfully generalized to classical systems in nonequilibrium steady states. Different theoretical formulations have been put forward \cite{aga72,cug94,Spe06,che08,Pro09,Bai09}, based, for instance, on the Fokker-Planck equation \cite{aga72}, the overdamped Langevin dynamics \cite{cug94,Spe06,che08}, the Hatano-Sasa fluctuation theorem \cite{Pro09}, or the dynamical activity \cite{Bai09}. Some of these modified fluctuation-response relations have been verified experimentally  using colloidal particles in a toroidal optical trap \cite{bli07,gom09,Bec10,Cil13}. 

Recently, Seifert and Speck have introduced a classification of steady-state fluctuation-dissipation theorems in the framework of stochastic thermodynamics, thus rationalizing previous methods that lead to apparently different results \cite{Sei10} (see also Ref.~\cite{Bai12}). Using a classical master equation approach, they have identified three main equivalence classes: the first variant contains a correlation function that involves no time derivatives (only functions of the steady-state distribution), the second variant is the unique form expressed in terms of time derivatives (of the stochastic entropy), whereas the last variant 
is the only one not requiring the {explicit knowledge of the} steady-state distribution. Infinitely many alternatives may be constructed via normalized linear combinations of the latter.  All these variants yield the same response and are thus equivalent. However,  the existence of different types of fluctuation-dissipation relations offers significant theoretical and experimental advantages. Theoretically, one kind of fluctuation-response theorem is usually easier to compute than the others, depending on the  concrete application. At the same time, the choice of the form crucially affects the accuracy of the experimental determination of the nonequilibrium response function, as shown in Ref.~\cite{Bec10}. Few attempts to extend  steady-state fluctuation-dissipation theorems to  open quantum systems have been presented \cite{wei71,Che12,San17}. However, a complete and unified picture is missing.

In this paper, we derive equivalence classes for generalized steady-state fluctuation-dissipation relations for open quantum systems using Markovian quantum Liouville equations. We discuss   the role of the noncommutativity of quantum operators and establish links to the Kubo response theory for closed quantum systems with unitary dynamics. To illustrate our unifying formalism, we  analytically compute the response function to a step perturbation for an open quantum system consisting of two weakly coupled  harmonic oscillators, each interacting with a  bath at a different temperature. This model describes the  coupling of a radiation mode with a vibrational mode in cavity optomechanics \cite{Mar14}, of two Bose condensates in bosonic Josephson junctions \cite{Jav86}, as well as the weak coupling limit of the Dicke Hamiltonian \cite{bra05}. The stationary state  corresponds to an equilibrium state when the two temperatures are equal and to a nonequilibrium steady state when they  are not. We use this model to stress the difference between classical and quantum responses, as well as between equilibrium and steady-state responses. Remarkably, we show that the equilibrium response function vanishes for  perturbations  that commute with the unperturbed Hamiltonian, whereas the  steady-state response  function does not. This underlines the profound disparity  between equilibrium and nonequilibrium quantum response theories.

 \textit{Quantum equivalence classes.}
We begin by deriving general equivalence classes for  steady-state quantum response functions. We consider open quantum systems with Hamiltonians of the form  {${H}(t) = {H}_0 + \varepsilon(t) {H}_I$}, where ${H}_0$ is the unperturbed Hamiltonian  and  $\varepsilon(t){H}_I$ the Hamiltonian perturbation with small time-dependent parameter $\varepsilon(t)$. We describe the open dynamics of the systems with density operator $\rho(t)$ using the Markovian quantum Liouville equation $d_t \rho(t) = \mathcal{L} \rho(t)$ \cite{Zwa01}. We expand the Liouville superoperator $\mathcal{L}$ to first order in $\varepsilon(t)$ as  $\mathcal{L} = \mathcal{L}_0 + \varepsilon(t) \mathcal{L}_1 + O(\varepsilon^2)$. The dynamics is thus  separated into an unperturbed part, $\mathcal{L}_0\cdot = -(i/\hbar) [{H}_0, \cdot] + \mathcal{D}[\cdot]$, where  $\mathcal{D}$ is the dissipator induced by the nonunitary coupling to the bath \cite{Bre02}, and a {perturbed}  part $\varepsilon(t) \mathcal{L}_1 \cdot= -(i\varepsilon(t) /\hbar) [{H}_I,\cdot]$. For fixed $\varepsilon$, we assume the existence of a stationary state ${\pi}_\varepsilon = {\pi}_0 + \varepsilon {\pi}_1$  such that $\mathcal{L} {\pi}_\varepsilon=0$. Starting with the unperturbed steady state ${\pi}_0$, the linear response of the density operator ${\rho}(t) = {\pi}_0+{\rho}_1(t)$ to a time-dependent perturbation $\varepsilon(t)$ may be written as
${\rho}_1(t) = \int_0^t ds\,\varepsilon(s)e^{\mathcal{L}_0(t-s)}\mathcal{L}_1 {\pi}_0$,
where $e^{\mathcal{L}_0(t-s)}$ is the evolution superoperator of the unperturbed system ${H}_0$ \cite{Zwa01}. The response of any observable of interest $A$ may then be calculated to linear order as,
\begin{eqnarray}
\langle {A}\rangle_\varepsilon(t) &=&  \langle {A} \rangle + \int_0^t ds \, \varepsilon(s) \text{Tr}\left\{{A} e^{\mathcal{L}_0(t-s)} \mathcal{L}_1 {\pi}_0   \right\}\nonumber \\
&=& \langle {A} \rangle+ \int_0^t ds \,\varepsilon(s) \mathcal{R}(t-s). 
\label{linearresp}
\end{eqnarray}
Here $\langle {A}\rangle_\varepsilon(t)= \text{Tr} \{A \rho(t)\}$ denotes the perturbed  expectation value of $A$ and $\langle {A}\rangle=\text{Tr} \{A \pi_0\}$  the corresponding unperturbed  expectation value. The response function is given by $\mathcal{R}(\tau)=\text{Tr}\left\{{A} e^{\mathcal{L}_0\tau} \mathcal{L}_1 {\pi}_0   \right\}$ with $\tau=t-s$. Equation  \eqref{linearresp} provides the basis for our quantum extension of the three equivalence classes identified in  Ref.~\cite{Sei10}.

\textit{Class one.} The first form $\mathcal{R}_
1(\tau)$ of the quantum response function is expressed as a correlation function with an observable ${B}_1=  (\mathcal{L}_1 {\pi}_0)/{\pi}_0$. It follows  from Eq.~\eqref{linearresp} by introducing the adjoint time evolution of the unperturbed dynamics ${A}(\tau) = {A} e^{\mathcal{L}_0 \tau}$ \cite{Bre02} and reads,
\begin{equation}
\mathcal{R}_
1(\tau) =   \langle A(\tau) B_1\rangle =\left\langle { A}(\tau)(\mathcal{L}_1 {\pi}_0)/\pi_0 \right\rangle.
\label{allgform}
\end{equation}
Expression \eqref{allgform} is a quantum generalization of the response function derived in Ref.~\cite{aga72} and is often referred to as Agarwal-form for this reason \cite{Sei10,Bai12}. This form readily shows that for a thermal stationary distribution, ${\pi}_0 = \exp(-\beta {H}_0)/Z_0$, with $Z_0$ the partition function, the quantum response vanishes when the perturbation commutes with the unperturbed Hamiltonian,   $[{H}_I,{H}_0]= 0$. This is not necessarily the case for a quantum nonequilibrium steady state, as we will discuss in detail below. This variant of the fluctuation-response theorem is distinguished by the fact that it contains only state variables and no time derivatives. Its drawback is that the observable $B_1$ involves the stationary distribution $\pi_0$, which is not always explicitly known in concrete situations.

\textit{Class two.} In the classical regime, the second variant is written in terms of  the time derivative of the $\varepsilon$-derivative of the stochastic entropy  of the system, $\partial_\varepsilon {S}_\varepsilon |_0 = - \partial_\varepsilon\ln  \pi_\varepsilon |_0 = -\pi_1/\pi_0$, along single trajectories  \cite{aga72,Sei10}. We obtain the second form $\mathcal{R}_2(\tau)$ of the quantum response function  by noting that  the stationary state of the Liouvillian has to vanish at every order, $(\mathcal{L}_0 + \varepsilon \mathcal{L}_1) ({\pi}_0 + \varepsilon {\pi}_1) = \mathcal{L}_0 {\pi}_0 + \varepsilon (\mathcal{L}_1 {\pi}_0 + \mathcal{L}_0 {\pi}_1) + O(\varepsilon^2)=0$. As a result, $\mathcal{L}_0 {\pi}_0= 0$ and $\mathcal{L}_1 {\pi}_0 =- \mathcal{L}_0 {\pi}_1$. Since  $\mathcal{L}_0$ is the generator of the open dynamics, we have  $e^{\mathcal{L}_0(t-s)}\mathcal{L}_0\rho=d_t e^{\mathcal{L}_0(t-s)}\rho =d_\tau e^{\mathcal{L}_0\tau}\rho=-d_s e^{\mathcal{L}_0(t-s)}\rho$. The response function  in Eq.~\eqref{linearresp} may therefore be rewritten as,
\begin{equation}
\begin{aligned}
\mathcal{R}_2(\tau)& = -\text{Tr}\left\{ {A}(\tau)\mathcal{L}_0  {\pi}_1\right\}=  -d_{\tau} \left\langle   {A}(\tau)  {\pi}_1/\pi_0\right\rangle,
\end{aligned}
\label{2formkuboref}
\end{equation}
In the limit of closed quantum systems at equilibrium, Eq.~\eqref{2formkuboref} reduces to the Kubo quantum response function, $\mathcal{R}_\text{K1}(\tau) = -\beta d_\tau\langle {A}(\tau) \tilde{{H}}_I\rangle$, since ${\pi}_1/{\pi}_0 = \beta \tilde{{H}}_I= \int_0^\beta d\lambda e^{-\lambda {H}_0} H_I e^{\lambda {H}_0}$ is the Kubo transform of  $H_I$ \cite{Kub57}.  The advantage of the Kubo transformation is that it allows to formulate classical and quantum equilibrium response functions in the same form by simply replacing an operator by its corresponding transform. Such a procedure can be carried over to steady-state response functions. In order to bring Eq.~\eqref{2formkuboref} in a form similar to the classical case, we first introduce a  generalized Kubo transformation:  Using the identity  $\partial_\varepsilon \pi_\varepsilon = \int_0^1 \pi_\varepsilon^\lambda (\partial_\varepsilon \ln \pi_\varepsilon) \pi_\varepsilon^{1-\lambda} d\lambda$, which holds for any density operator \cite{Ama00},  we obtain, for $\varepsilon = 0$, $\partial_\varepsilon {\pi}_\varepsilon|_0/{\pi}_0 = \int_0^1 {\pi}_0^\lambda (\partial_\varepsilon \ln {\pi}_\varepsilon)|_0{\pi}_0^{-\lambda}d\lambda \eqqcolon \overline{\partial_\varepsilon \ln {\pi}_\varepsilon |_0}$, where we have defined the  transform $\overline{\partial_\varepsilon \ln {\pi}_\varepsilon|_0}$ of ${\partial_\varepsilon \ln {\pi}_\varepsilon}|_0$ . The latter reduces to the usual Kubo transform for a thermal state, $\pi_\varepsilon =  e^{-\beta (H_0 + \varepsilon H_I)}/Z_\varepsilon$. We accordingly find, 
\begin{equation}
\mathcal{R}_2(\tau) =   -{d_\tau} \left\langle   {A}(\tau)\overline{\partial_\varepsilon \ln {\pi}_\varepsilon|_0} \right\rangle  =  {d_\tau} \left\langle   {A}(\tau)\overline{\partial_\varepsilon {S}_\varepsilon |_0}\right\rangle ,
\label{2formalt}
\end{equation}
where we have introduced the quantum analog of the stochastic entropy ${S}_\varepsilon = - \ln {\pi}_\varepsilon$. Noting furthermore that two-time correlation functions for open quantum systems are defined as $\langle A(t)B(s)\rangle= \text{Tr}\left\{ A e^{\mathcal{L}_0(t-s)} B e^{\mathcal{L}_0 s}\rho(0)\right\}$ \cite{Bre02,Car93}, we obtain, with $\rho(0) = \pi_0$ and  $d_s e^{\mathcal{L}_0 s} \pi_0=0$,
\begin{equation}
\mathcal{R}_2(\tau) = - d_s \left\langle   {A}(\tau)\overline{\partial_\varepsilon {S}_\varepsilon|_0 } \right\rangle = -\left\langle   {A}(t)d_s\overline{\partial_\varepsilon {S}_\varepsilon(s)|_0} \right\rangle .
\label{2form}
\end{equation}
Formula \eqref{2form} is a quantum extension of the response function of Refs.~\cite{aga72,Sei10}. The results  presented in Refs.~\cite{Che12,San17} are related to this variant \cite{tobe}. It can be formally written in Liouville space as a correlation function with the observable $B_2(s)= -d_s \overline{\partial_\varepsilon {S}_\varepsilon(s) |_0}$.
In general, for noncommuting operators, {$\partial_\varepsilon \ln {(\pi_0 +\varepsilon\pi_1)} |_0 \neq{\pi}_1/{\pi}_0$}, implying that $\overline{\partial_\varepsilon {S}_\varepsilon} \neq {\partial_\varepsilon {S}_\varepsilon}$.  Consequently, the quantum response function \eqref{2form} cannot be  written in terms of the stochastic entropy, $ - \left\langle   {A}(t)d_s\partial_\varepsilon {S}_\varepsilon(s) |_0\right\rangle $, as in the classical limit, unless $[{\pi}_0,{\pi}_1] = 0$ \cite{sm}.
The variant \eqref{2form} is the only one where the  response function is given as a correlation function with a time derivative of a state variable, namely the  formal time derivative of the generalized Kubo transformed $\varepsilon$-derivative of the stochastic entropy, $\overline{\partial_\varepsilon {S}_\varepsilon|_0}$. 

Let us additionally mention that there is an alternative way of writing the quantum response function \eqref{2form} without using any correlation function. We indeed have,
\begin{equation}
\mathcal{R}_\text{2,alt}(\tau) = - {d_\tau}{\partial_\varepsilon}\text{Tr}\left\{  { A}(\tau)  {\pi}_\epsilon \right\}\vline_{0} = - {d_\tau} {\partial_\varepsilon}\left\langle  {A}(\tau) \right\rangle_\varepsilon \vline_{0},
\label{version2alt}
\end{equation}
where $\langle {A}(\tau)\rangle_\varepsilon = \text{Tr}\{A(\tau) \pi_\varepsilon\}\neq \langle {A}\rangle_\varepsilon(\tau)$ is the perturbed expectation value of the observable ${A}(\tau)={A} e^{\mathcal{L}_0 \tau} $ evolved via the unperturbed dynamics. This form offers an intuitive interpretation of dynamic response theory: at any fixed time, ${\partial_\varepsilon}\left\langle  {A}(\tau) \right\rangle_\varepsilon \vline_{\varepsilon = 0}$ can be seen as the static susceptibility, that is, the static response of the system to the external perturbation \cite{Zwa01}. The dynamic response function \eqref{version2alt} then follows as the time derivative of the time-dependent susceptibility. This form often enables a simple evaluation of the response function (see below).

\textit{Class three.} Classically, the third form is the unique one  that does not  explicitly involve the stationary distribution  \cite{Bai09,Sei10}. This type of fluctuation-response relation is therefore of  advantage when the steady-state distribution is not specifically known. Within our quantum Hamiltonian perturbation approach, such a variant may be derived from   Eq.~\eqref{allgform} by realizing that  $\mathcal{L}_1 \cdot= -(i/\hbar)[{H}_I,\cdot]$. We then obtain,  
\begin{equation}
\mathcal{R}_3 (\tau) =  \frac{i}{\hbar} \text{Tr}\left\{{\pi}_0 \left[{H}_I,{A}(\tau)\right] \right\}= - \left\langle \mathcal{L}_1 {A}(\tau)\right\rangle.
\label{Kommform}
\end{equation}
In contrast to Eqs.~\eqref{allgform}-\eqref{version2alt}, the  response function  \eqref{Kommform} is given as an expectation value of operators that do not explicitly depend on either $\pi_0$, $\pi_1$ or $\pi_\varepsilon$ (see also Ref.~\cite{wei71} for an alternative approach). In the limit of unitary quantum systems at equilibrium, Eq.~\eqref{Kommform} reduces to the Kubo quantum response function $\mathcal{R}_\text{K2}(\tau) = (i/\hbar) \langle [{H}_I,{A}(\tau)]\rangle$  \cite{Kub57}. Interestingly, expression \eqref{Kommform} indicates that the quantum response function vanishes when the time evolved observable  $A(\tau)$ commutes with the perturbation Hamiltonian $H_I$. This form further shows that the quantum response function is  equal to the expectation value of the imaginary part of the correlation function $\langle {A}(\tau) {H}_I \rangle = \langle {A}(\tau) {H}_I + {H}_I {A}(\tau)\rangle/2 + i \langle [{A}(\tau),{H}_I]/i\rangle/2$  for an nonequilibrium steady state, as in the unitary limit \cite{Kub57}.

\begin{figure}[t]
\includegraphics[width=0.37\textwidth]{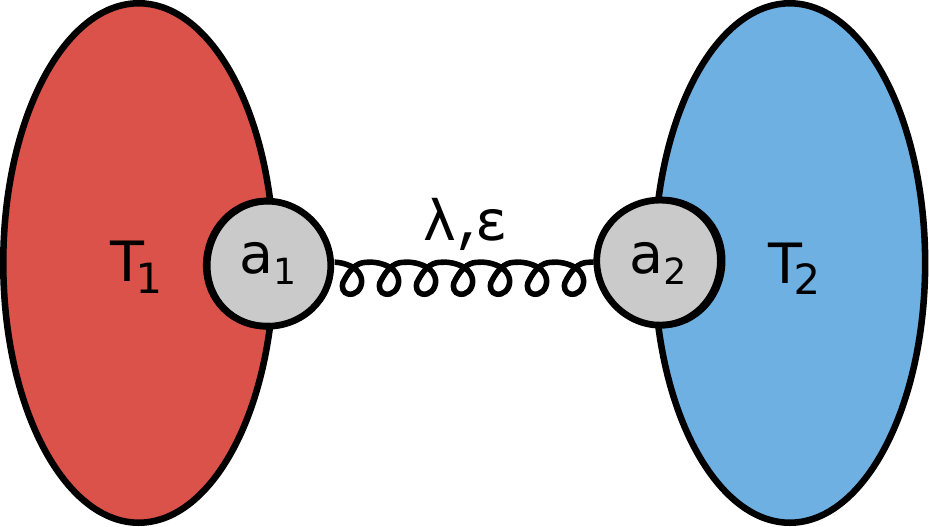}
\caption{Steady-state model. Two quantum harmonic oscillators are weakly coupled to each other with coupling strength $\lambda$. Each of them interacts with a bath with a different temperature $T_j$, $(j=1,2)$. A nonequilibrium steady state is established when the two temperatures are different and heat flows from one oscillator to the other. We examine the response of oscillator 1  when the coupling is modulated by $\varepsilon(t)$.} 
\label{setup}
\end{figure}

\textit{Example.} Our results are applicable to general open quantum systems. As an illustration, we now consider a system consisting of two weakly coupled harmonic oscillators, each interacting with its own reservoir at a different temperature (Fig.~1). By properly tuning the parameters of the system, this model   allows one to compare different response regimes: unitary/dissipative, equilibrium/steady state and classical/quantum. In particular, a nonequilibrium steady state is established when the two  bath temperatures are different. The (unperturbed) Hamiltonian of the system is \cite{Mar14,Jav86,bra05},
\begin{equation}
{H}_0 = \hbar \omega_1 {a}_1^\dagger{ a}_1 +  \hbar (\omega_1+\delta){ a}_2^\dagger {a}_2 + \hbar \lambda ({a}_1 {a}_2^\dagger +{ a}_1^\dagger {a}_2),
\end{equation}
where (${a}_j^\dagger,{ a}_j$) ($j=1, 2$) are the ladder operators of the  oscillators with respective frequencies $\omega_1$ and $\omega_2=\omega_1+ \delta$ (with detuning $\delta$). The coupling parameter is denoted by $\lambda$.
\begin{figure}[t]
\includegraphics[width=0.48\textwidth]{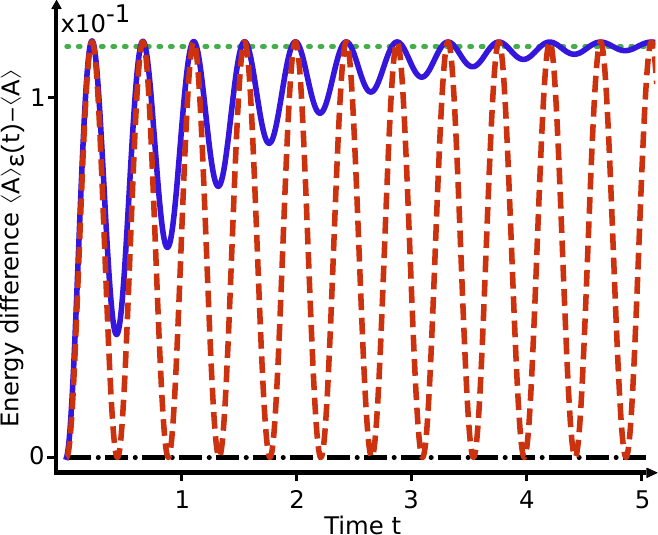}
\caption{Quantum response of the (dimensionless) energy of the first oscillator, $\langle{A}\rangle_\varepsilon(t)-\langle A\rangle$, with $A = \beta_1\hbar \omega_1 {a}_1^\dagger {a}_1$, to a step perturbation, $\varepsilon(t) = \varepsilon \Theta(t)$, of the coupling between the two harmonic oscillators. The steady-state response $(\lambda \neq 0)$ (blue solid), Eq.~\eqref{responsefunctionbsp}, asymptotically approaches the perturbed value (green dotted). By contrast, the equilibrium response $(\lambda = 0)$ (black dotted-dashed) vanishes and the unitary response $(\gamma=0$) (red dashed), Eq.~\eqref{unit}, keeps oscillating and fails to reach the perturbed value of the observable $A$. Parameters are $\omega_1= 2.4$, $\delta= 10.1$, $\gamma= 0.7$, $\lambda= 5$, $\varepsilon= 0.11$, $\beta_1= 0.092$ and $\beta_2= 0.0008$.}
\label{f1}
\end{figure}
The (unperturbed)  Liouville superoperator reads \cite{Cam16},
\begin{equation}
\mathcal{L}_0 \cdot= - (i/\hbar) [{H}_0, \cdot] + \sum_{j=1}^2\mathcal{D}_j[\cdot],
\end{equation}
with the two nonunitary dissipators $\mathcal{D}_j$ induced by the interaction with the heat reservoirs,
\begin{eqnarray}
\mathcal{D}_j[{\rho}] &=& \gamma (n_j + 1) \left [{a}_j {\rho} {a}_j^\dagger - \frac{1}{2}\left ({a}_j^\dagger {a}_j {\rho} + {\rho} {a}_j^\dagger {a}_j \right)\right] \nonumber \\
&+& \gamma n_j  \left[ {a}_j^\dagger {\rho} {a}_j +  \frac{1}{2}\left({a}_j {a}_j^\dagger  {\rho} + {\rho} {a}_j {a}_j^\dagger \right)\right] .
\end{eqnarray}
Here  $n_j= [\exp(\beta_j\omega_j)-1]^{-1}$ is the thermal occupation number at inverse temperature $\beta_j$ and $\gamma$  is the damping constant. For concreteness, we apply a step perturbation $\varepsilon(t) {H}_I = \hbar \varepsilon(t) ({a}_1 {a}_2^\dagger + {a}_1^\dagger {a}_2)$, with $\varepsilon(t) = \varepsilon \Theta(t)$, to the coupling between the  harmonic oscillators and look at the response of the (dimensionless) energy of the first oscillator, $A = \beta_1\hbar \omega_1 {a}_1^\dagger {a}_1$. We note that the unperturbed system is in a thermal state for $\lambda =0$ and in a nonequilibrium steady state for $\lambda \neq 0$. The two quantum oscillators are moreover closed with unitary dynamics in the absence of damping, $\gamma =0$. Finally, the classical regime is achieved in the high-temperature limit $\beta_j\hbar \omega_j \ll1$.

We determine the quantum response function using the forms \eqref{version2alt} and \eqref{Kommform}. To this end, we   first evaluate the steady-state density operator of the  system and then calculate the time dependence of the observable $A$. Due to the quadratic nature of the Hamiltonian, it is convenient to employ the Gaussian characteristic function $\chi(x_1,p_1,x_2,p_2) = \exp(i \vec P \vec {\bar y}  -  \vec P^T \bar \sigma \vec P/2)$ of the system, with coordinate vector $\vec P = (x_1,p_1,x_2,p_2)^T$,  symplectic mean vector $\vec{\bar{y}}$ and covariance matrix $\bar\sigma$ to compute the stationary distribution \cite{Cam16}. The time evolution of  $A$ is obtained via matrix exponentiation. We find \cite{sm},
\begin{equation}
\begin{aligned}
\mathcal{R}_3(\tau) &= i\beta_1\hbar \omega_1 \langle [{a}_1 {a}_2^\dagger + {a}_1^\dagger {a}_2, {a}_1^\dagger {a}_1(\tau)]\rangle\\
&=  e^{-\gamma \tau} \frac{\gamma(\delta^2 + 4\lambda^2 \cos z\tau) + (\gamma^2 + \delta^2) z \sin z\tau}{z^2 (\gamma^2 + z^2)(2\lambda\Delta n\beta_1\hbar \omega_1)^{-1} }\\
&= - \beta_1\hbar \omega_1{d_\tau}{\partial_\varepsilon}\left\langle {a}^\dagger {a}(\tau) \right\rangle|_{\varepsilon=0} = \mathcal{R}_\text{2,alt}(\tau),
\end{aligned}
\label{responsefunctionbsp}
\end{equation}
where $\Delta n = n_2-n_1$ and $z = \sqrt{\delta^2 + 4 \lambda^2}$. The two  different forms $\mathcal{R}_\text{2,alt}(\tau)$ and $\mathcal{R}_3(\tau)$ thus yield the same result, as expected. However, this is not obvious from their definitions \eqref{version2alt} and \eqref{Kommform}, since $\mathcal{R}_\text{2,alt}(\tau)$ displays a $\varepsilon$-dependence, while $\mathcal{R}_3(\tau)$ does not.

\begin{figure}[t]
\includegraphics[width=0.45\textwidth]{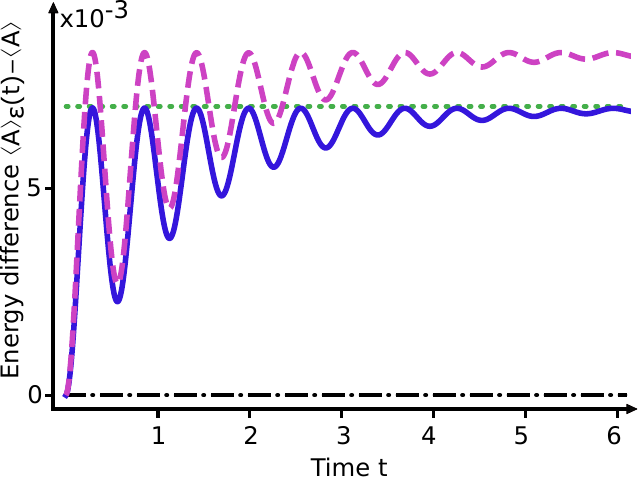}
\caption{Steady-state response of the (dimensionless) energy of the first oscillator, $\langle{A}\rangle_\varepsilon(t)-\langle A\rangle$, with $A = \beta_1\hbar \omega_1 {a}_1^\dagger {a}_1$, to a step perturbation, $\varepsilon(t) = \varepsilon \Theta(t)$, of the coupling between the two harmonic oscillators. The quantum  response {$(\beta_1\hbar \omega_1 \gg1)$} (blue solid), Eq.~\eqref{responsefunctionbsp}, asymptotically approaches the perturbed value (green dotted). By contrast, the classical response {$(\beta_1\hbar \omega_1 \ll1)$} (purple dashed), Eq.~\eqref{classicallim} although proportional to the quantum response fails to reach the perturbed value of the observable $A$. Parameters are $\omega_1= 2.4$, $\delta= 10.1$, $\gamma= 0.7$, $\lambda= 2.3$, $\varepsilon= 0.11$, $\beta_1= 0.164$ and $\beta_2= 0.416$.}
\label{f2}
\end{figure}

Three different response regimes may be distinguished, as can be seen in  Figs. 2 and 3 that represent the response difference, $\langle{A}\rangle_\varepsilon(t)-\langle A\rangle$ for various system parameters: (i) In the thermal limit $\lambda \rightarrow 0$, the unperturbed quantum oscillator is in an equilibrium state and the quantum response function  \eqref{responsefunctionbsp} vanishes (black dotted-dashed line). By contrast, the steady-state response is different from zero (blue solid line) and approaches the perturbed stationary value at large times (green dotted line). This example emphasizes the fundamental difference between equilibrium and steady-state quantum response theories. The response function can actually be used to distinguish thermal and nonthermal states via their different behavior for a perturbation that commutes with $H_0$ for $\delta=0$ \cite{sm}. (ii) In the unitary limit $\gamma=0$, when the interaction with the two heat reservoirs is switched off, the quantum response function \eqref{responsefunctionbsp} reduces to,
\begin{equation}
\label{unit}
\mathcal{R}_\text{unitary}(\tau) = 2 \lambda\Delta n \beta_1 \hbar\omega_1 \frac{\delta^2}{z^{3}}\sin z\tau.
\end{equation}
We observe (Fig.~2) that the perturbed observable (red dashed line) exhibits in this situation oscillations with the same oscillation period as in the nonunitary case ($\gamma\neq 0$) (blue solid line). However, it never reaches its  perturbed value (green dotted line) due to the absence of external damping. (iii) Finally, in the classical limit, $\beta_j\hbar \omega_j \ll1$, the Bose distribution reduces to the Boltzmann distribution and the response function \eqref{responsefunctionbsp} simplifies to,
\begin{equation}
\begin{aligned}
\mathcal{R}_\text{classical}(\tau) &=  \frac{\beta_1\omega_1-\beta_2\omega_2}{\beta_2\omega_2}  \frac{\mathcal{R}_3(\tau)}{\Delta n \beta_1\hbar\omega_1}.
\end{aligned}
\label{classicallim}
\end{equation} 
The classical response function \eqref{classicallim} is hence proportional to the quantum response function \eqref{responsefunctionbsp}. However, it predicts the wrong perturbed value of the observable $A$ (purple dotted line in Fig.~3), stressing the difference between classical and quantum response theories. 

\textit{Conclusions.}
We have performed an extensive study of the nonequilibrium steady-state response of  open quantum systems described by Markovian Liouville equations. We have concretely derived  quantum extensions of the equivalence classes for classical response functions introduced in Ref.~\cite{Sei10}. We have for each of them analyzed  the role of noncommuting operators and identified conditions under which the quantum response vanishes when some operators commute. We have further shown that the second quantum form cannot be written in terms of the stochastic entropy of the system, as in the classical case, but instead in terms of the generalized Kubo transform of the latter. We have additionally established that the second and third quantum variants are nonunitary extensions of the familiar Kubo response functions to which they reduce in the limit of closed quantum systems at equilibrium. We have finally illustrated our results with  an analytically solvable model of two weakly coupled open quantum harmonic oscillators and compared various response regimes including  unitary/dissipative, equilibrium/steady state, and classical/quantum limits. We have shown, in particular, that the equilibrium quantum response can vanish in instances where the steady-state quantum response does not. Our findings not only provide a unified picture of nonequilibrium quantum response theory, they also offer different, but equivalent, approaches to evaluate  steady-state response functions, depending on the specific problem considered.

\textit{Acknowledgments.}
We  thank  Tobias Donner for  discussions and acknowledge  financial support from the DFG (Contract No FOR 2724).

\newpage
\section{Supplemental Material}
\textit{Properties of the second from.} 
We here show that the  response function  $R_2(\tau)$, Eq.~\eqref{2form}, vanishes for a thermal state when $[\pi_0,\pi_1] = 0$. For commuting distributions, we indeed have  $\partial_\varepsilon \ln(\pi_0 + \varepsilon \pi_1)|_0 = \pi_1/\pi_0$. Moreover, for a thermal state $\pi_\varepsilon \propto e^{-\beta(H_0 + \varepsilon H_I)}$, the operator $\pi_1$ can be given explicitly via the Kubo transform, $\pi_1 \propto  \tilde{H}_I \pi_0$ (see Eq.~(7.12) in Ref.~\cite{Zwa01}). As a result, the condition $[\pi_0,\pi_1] = 0$ implies that  $[H_0,H_I]=0$ since the Kubo transform $\beta\tilde{H}_I = \int_0^\beta d\lambda e^{-\lambda H_0} H_I e^{\lambda H_0}$ consists of exponentials that are proportional to $H_0$. Equation  \eqref{allgform} then shows that $\mathcal{L}_1 \pi_0 = 0$ and that the response thus vanishes.

{In the unitary limit, the formal identity in Liouville space for the correlation function \eqref{2form} becomes one at the operator level in Hilbert space. In that case,  the unitary time-evolution operators $A  e^{\mathcal{L}_0t}= U^\dagger(t,0) A U(t,0) = A(t)$ can be used to obtain the Heisenberg representation of the generalized stochastic entropy $\left\{ U^{\dagger}(t,s) A U(t,s) B U(s,0)\rho(0)U^\dagger(s,0)\right\} = \langle A(t) B(s)\rangle$. The latter equality  gives the observable $- d_s \overline{\partial_\varepsilon {S}_\varepsilon|_0(s)}$ a meaning in Hilbert space.}

\textit{Calculations for the coupled-oscillator model.}
We first determine the steady-state solution by introducing following Ref.~\cite{Cam16}
 the symmetric characteristic function $\chi (\alpha_1,\alpha_2)= \langle {D}_1(\alpha_1)\otimes {D}_2(\alpha_2) \rangle$, where ${D}_i(\alpha_i) = \exp( \alpha_i {a}_i^\dagger -  \alpha_i^* {a}_i )$ is  the displacement operator. The {symmetric} moments are directly obtained by differentiation,
 \begin{equation}
\langle{a}_i^{\dagger k} {a}_j^l\rangle_s = \frac{d^k}{d\alpha_i^k} \frac{d^l}{(-\alpha_j^*)^l} \chi(\alpha_1,\alpha_2)|_{\alpha_1=\alpha_2 =0},
\label{symplectrafo}
\end{equation} 
{where $\langle\cdot\rangle_s$ is the expectation value of the symmetrized version of the operators $a_i^{\dagger k} a_j^l$.}
The  equation for the characteristic function can be derived from the quantum Liouville equation by using,
\begin{equation}
\frac{d}{dt}\chi(\alpha_1,\alpha_2) = \text{Tr}\{  {D}_1(\alpha_1)\otimes {D}_2(\alpha_2) \dot{\rho} \},
\end{equation}
together with the identities,
\begin{eqnarray}
{D}_i {a}_i^\dagger &=& \left(- \frac{\alpha_i^*}{2} + \frac{d}{d\alpha_i} \right){D}_i, \,{D}_i {a}_i = \left(- \frac{\alpha_i}{2}- \frac{d}{d\alpha_i^*} \right){D}_i, \nonumber \\
{a}_i^\dagger {D}_i  &=& \left( \frac{\alpha_i^*}{2} + \frac{d}{d\alpha_i} \right){D}_i, \,{a}_i {D}_i = \left( \frac{\alpha_i}{2} - \frac{d}{d\alpha_i^*} \right){D}_i.
\end{eqnarray}
We then obtain the differential equation,
 \begin{eqnarray}
\frac{d\chi}{dt} &=& \left\{ \sum_{j=1}^2\left[ \omega_j \left(x_j\frac{d}{dp_j} - p_j \frac{d}{dx_j}\right) \right. \right. \nonumber\\
&-& \left. \frac{\gamma_j}{2}(2n_j + 1)\left(x_j^2 + p_j^2\right) - \left.\frac{\gamma_j}{2}\left(x_j \frac{d}{dx_j} + p_j \frac{d}{dp_j}\right) \right] \right. \nonumber \\
&+&\left.  \lambda \left(x_2 \frac{d}{dp_1} -p_2 \frac{d}{dx_1}+x_1 \frac{d}{dp_2} -p_1 \frac{d}{dx_2}\right)                \right\}\chi, 
\label{chardgl}
\end{eqnarray}
with  ${x}_i = ({a}_i^\dagger + {a}_i)/\sqrt{2}$ and ${p}_i = i( {a}_i-{a}_i^\dagger)/\sqrt{2}$, as well as $\alpha_i = x_i + i p_i$ and $d/d\alpha_i = (d/dx_i - i d/dp_i)/2$. Equation \eqref{chardgl} can be 
 be solved  by using the Gaussian ansatz $\chi(x_1,p_1,x_2,p_2)  = \exp(i \vec P \vec{\bar{y}} -  \vec P^T \bar \sigma \vec P/2)$ with $\vec P = (x_1,p_1,x_2,p_2)^T$ and $\vec{\bar{y}} = (\bar{y}_1,\bar{z}_1,\bar{y}_2,\bar{z}_2)^T$. {Using Eq.~\eqref{symplectrafo},} the variables $(\bar{y}_i,\bar{z}_i)$ and the matrix $\bar \sigma$ can be related to the average quadratures and the covariance matrix $\sigma$ via,
 \begin{eqnarray}
\langle {x}_i\rangle &=& \frac{\bar{z}_i}{\sqrt{2}}, \quad  \langle {p}_i\rangle = - \frac{\bar{y}_i}{\sqrt{2}}\\
\sigma_{x_i x_j} &=& \frac{1}{2}\bar{\sigma}_{p_i p_j},\, \sigma_{p_i p_j} = \frac{1}{2}\bar{\sigma}_{x_i x_j}, \, \sigma_{x_i p_j} = -\frac{1}{2}\bar{\sigma}_{x_j p_i}
\end{eqnarray}
The steady-state solution is given by vanishing first moments, $\langle {x}_{1,2}\rangle = \langle {p}_{1,2}\rangle = 0$,   and the covariance matrix,
\begin{widetext}
\begin{equation}
\sigma= \zeta
\left(
\begin{array}{cccc}
D +n_1 + \frac{1}{2} & 0& -\delta C  & -\gamma C\\
0 & D +n_1 + \frac{1}{2} &  \gamma C & -\delta C\\
-\delta C & \gamma C & D +n_2 + \frac{1}{2} & 0\\
-\gamma C & - \delta C & 0 & D +n_2 + \frac{1}{2}
\end{array}
\right),
\label{covmatrix}
\end{equation}
\end{widetext}
with the three parameters,
\begin{eqnarray}
\zeta &=& \frac{\gamma^2 + \delta^2}{4 \lambda^2 + \gamma ^2 + \delta^2},\\
 D &=& \frac{2 \lambda^2(n_1 + n_2 + 1)}{\gamma^2 + \delta^2}, \\
 C &=& \frac{\lambda(n_1 - n_2)}{\gamma^2 + \delta^2}.
\end{eqnarray}

The time evolution of the observable of interest $A$ is obtained by solving the adjoint master equation by finding  operators that form a closed set of linear differential equations  \cite{Bre02}. For the considered quadratic system, these operators are given by $\vec v(t)^T = \left({a}_1^\dagger {a}_1(t),a_1^2(t), a_1^{\dagger2}(t), a_2^\dagger a_2(t),a_2^2(t),a_2^{\dagger 2}(t),\right.$  $\left.a_1a_2(t),a_1a_2^\dagger(t),a_1^\dagger a_2(t),a_1^\dagger a_2^\dagger(t) \right)$. They obey the ensemble  of linear differential equations,
\begin{equation}
\frac{d}{dt}\vec v(t) = M \vec v (t)+\vec w. 
\end{equation}
with  $\vec w = \left(n_1\gamma,0,0,n_2 \gamma,0,0,0,0,0,0\right)$. The  matrix $M$ explicitly reads,
\begin{widetext}
\begin{equation}
M= \left(
\begin{array}{cccccccccc}
-\gamma &0&0&0&0&0&0& i\lambda&-i\lambda&0\\
0&- 2 i\omega_1-\gamma&0&0&0&0&- 2 i \lambda&0&0&0\\
0&0&2i\omega_1 -\gamma&0&0&0&0&0&0&2i\lambda\\
0&0&0&-\gamma&0&0&0&-i\lambda&i\lambda&0\\
0&0&0&0&-2i\omega_2-\gamma&0&-2i\lambda&0&0&0\\
0&0&0&0&0& 2 i \omega_2-\gamma&0&0&0& 2i\lambda\\
0&-i\lambda&0&0&-i\lambda&0&-i\omega_{12} -\gamma&0&0&0\\
i\lambda&0&0&-i\lambda&0&0&0&-i\Delta \omega -\gamma&0&0\\
-i\lambda&0&0&i\lambda&0&0&0&0&i\Delta\omega-\gamma&0\\
0&0&i\lambda&0&0&i\lambda&0&0&0&i\omega_{12}-\gamma
\end{array}
\right)
\end{equation}
\end{widetext}
with $\omega_{12} = \omega_1 + \omega_2$ and $\Delta \omega = \omega_1 -\omega_2$.  The time dependence of the number operator, $a_1^\dagger a_1$,  of the first oscillator   follows via  matrix exponentiation, 
\begin{eqnarray}
a_1^\dagger a_1(t) &=& f(t) a_1^\dagger {a}_1 + g(t) a_1^2 + h(t) a_1^{\dagger 2}+ j(t) a_2^\dagger a_2 \nonumber \\
&+&l(t)a_2^2 + m(t) a_2^{\dagger 2}+ n(t) {a}_1 a_2+ p(t) a_1 a_2^\dagger \nonumber\\
& +& q(t) a_1^\dagger a_2 + r(t) a_1^\dagger a_2 ^\dagger + s(t).
\label{adjointtevo}
\end{eqnarray}
The various functions appearing in Eq.~\eqref{adjointtevo} are given by,
\begin{eqnarray}
f(t) &=& e^{-\gamma t} \left(\delta ^2+2 \lambda ^2+2 \lambda ^2 \cos z t\right)/z^2,\\
j(t) &=& -2 \lambda^2 e^{-\gamma t} \left(\cos zt-1\right)/z^2,\\
p(t) &=&\lambda  e^{-\gamma t} \left(-\delta +i z \sin zt+\delta  \cos zt\right)/z^2,\\
q(t) &=& \lambda  e^{-\gamma t} \left(-\delta -i z \sin zt+\delta  \cos zt\right)/z^2,\\
r(t) &=& 0, ~g(t) =0,~ h(t) =0,\\
 l(t)&=&0,~m(t)=0,~n(t)=0, 
\end{eqnarray}
together with
\begin{eqnarray}
s(t) &=& \frac{1}{z^3 \left(\gamma^2+\delta ^2+4 \lambda ^2\right)} e^{-\gamma t}\nonumber \\
&\times&\left\{ z \left[\left(\gamma^2+z^2\right) \left(\delta ^2 n_1+2 \lambda ^2 (n_1+n_2)\right)\right.\right.\nonumber \\
&-&\left.z^2 e^{\gamma t} \left(n_1 \left(\gamma^2+\delta ^2\right)+2 \lambda ^2 (n_1 + n_2)\right)\right] \nonumber \\
&+&\left. 2 \gamma \lambda ^2 (n_1 - n_2) \left(\gamma z \cos zt -z^2 \sin zt \right)\right\}.
\end{eqnarray}
 For $\lambda = 0$, Eq.~\eqref{adjointtevo} simplifies to $a_1^\dagger a_1(t) = e^{-\gamma t} a_1^\dagger a_1 + (1-e^{-\gamma t})n_1$, as expected for a thermal  oscillator \cite{Bre02}.

\textit{Response  for the two-oscillator model for $\delta = 0$.}
We here explictly show that the steady-state response may be different from zero in cases where the equilibrium response vanishes.
For zero detuning, $\delta =0$, the perturbation commutes with the unperturbed Hamiltonian, $[H_0, H_I] = \hbar \omega_1[a_1^\dagger a_1 + a_2^\dagger a_2 + \lambda H_I , a_1^\dagger a_2 + a_1 a_2^\dagger]= 0$, implying that the response vanishes for a thermal state. However, in that limit the  response function \eqref{responsefunctionbsp} reads,
\begin{equation}
\mathcal{R}_{\delta = 0}(\tau) = e^{-\gamma \tau} \Delta n \gamma \frac{2 \lambda \cos(2\lambda \tau) + \gamma \sin(2\lambda \tau)}{(\gamma^2 + 4 \lambda ^2)(\beta_1\hbar \omega_1)^{-1}},
\label{delta0ex}
\end{equation}
which is in general nonzero. { The thermal response thus vanishes for $\lambda\neq 0$, while the steady-state response is finite. In the thermal limit, $\lambda\rightarrow 0$, Eq.~\eqref{delta0ex} clearly vanishes, as it should.}


\begin{thebibliography}{99}
  \bibitem{nyq28} H. Nyquist, Thermal Agitation of Electric Charge in Conductors, Phys. Rev. \textbf{32}, 110 (1928).
  \bibitem{cal51}  H. B. Callen and T. A. Welton, Irreversibility and Generalized Noise, Phys. Rev. \textbf{83}, 34 (1951).
  \bibitem{Kub57} R. Kubo, Statistical-mechanical theory of irreversible Processes. I. General theory and simple applications to magnetic and conduction problems, {J. Phys. Soc. Jpn.} \textbf{12}, 6 (1957).
  \bibitem{Kub66} R. Kubo, The fluctuation-dissipation theorem, Rep. Prog. Phys. \textbf{29} 255 (1966).
  \bibitem{han82} P. H\"anggi and H. Thomas, Stochastic processes: time-evolution, symmetries and linear response, Phys.
Rep. \textbf{88}, 207 (1982).
\bibitem{Zwa01} R. Zwanzig, \textit{Nonequilibrium Statistical Mechanics}, (Oxford University Press, Oxford, 2001).
\bibitem{mar08} U. M. B. Marconi, A. Puglisi, L. Rondoni, and A. Vulpiani, Fluctuation-Dissipation: Response Theory in Statistical Physics, Phys. Rep. \textbf{461}, 111 (2008).
\bibitem{aga72} G. S. Agarwal, Fluctuation-dissipation theorems for systems in non-thermal equilibrium and applications,
Z. Phys. \textbf{252}, 25 (1972).
\bibitem{cug94} L. Cugliandolo, J. Kurchan, and G. Parisi, Off equilibrium dynamics and aging in unfrustrated systems
J. Physique I \textbf{4}, 1641 (1994).
\bibitem{Spe06} T. Speck and U. Seifert, Restoring a fluctuation-dissipation theorem in a nonequilibrium steady state, Europhys. Lett. \textbf{74} (3) 391 (2006).
\bibitem{che08} R. Chetrite, G. Falkovich, and K. Gawedzki, Fluctuation relations in simple examples of non- equilibrium steady states, J. Stat. Mech.  \textbf{P08005} (2008). 
\bibitem{Pro09} J. Prost, J. Joanny, and J. M. R. Parrondo, Generalized fluctuation-dissipation theorem for steady state systems, Phys. Rev. Lett. \textbf{103}, 090601 (2009).
\bibitem{Bai09} M. Baiesi, C. Maes, and B. Wynants, Fluctuations and Response of Nonequilibrium States, Phys. Rev. Lett. \textbf{103}, 010602 (2009).
\bibitem{bli07} V. Blickle, T. Speck, C. Lutz, U. Seifert, and C. Bechinger, Einstein relation generalized to nonequilibrium,
Phys. Rev. Lett. \textbf{98}, 210601 (2007).
\bibitem{gom09} J. R. Gomez-Solano, A. Petrosyan, and S. Ciliberto, Experimental Verification of a Modified Fluctuation-Dissipation Relation
for a Micron-Sized Particle in a Nonequilibrium Steady State, Phys. Rev. Lett. \textbf{106}, 200602 (2009).
\bibitem{Bec10} J. Mehl, V. Blickle, U. Seifert, and C. Bechinger, Experimental accessibility of generalized fluctuation-dissipation relations for nonequilibrium steady states, Phys. Rev. E \textbf{82}, 032401 (2010).
\bibitem{Cil13} S. Ciliberto, R. Gomez-Solano, and A. Petrosyan, Fluctuations, linear response, and currents in out-of-equilibrium systems, Annu. Rev. Cond. Mat. Phys. \textbf{4}, 235 (2013).
\bibitem{Sei10} U. Seifert and T. Speck, Fluctuation-dissipation theorem in nonequilibrium steady states, {EPL} \textbf{89},  10007, (2010).
\bibitem{Bai12} M. Baiesi and C. Maes, An update on the nonequilibrium linear response, New J.  Phys. \textbf{15},  013004 (2013).
%\bibitem{Bec09} V. Blickle, J. Mehl, C. Bechinger, Relaxation of a colloidal particle into a nonequilibrium steady state, PRE \textbf{79}, 6 (2009).
\bibitem{wei71} W. Weidlich, Fluctuation-dissipation theorem for a class of stationary open systems, Z. Phys. \textbf{248}, 234 (1971).
\bibitem{Che12} R. Chetrite and K. Mallick, Quantum fluctuation relations for the Lindblad master equation, J. Stat. Phys. \textbf{148}, 480 (2012).
\bibitem{San17} M. Mehboudi, A. Sanpera, and J. M. R.  Parrondo, Generalized fluctuation-dissipation relation for quantum Markovian systems, Quantum \textbf{2}, 66 (2018).
\bibitem{Mar14} M. Aspelmeyer, T. J. Kippenberg, and F. Marquardt, Cavity optomechanics, Rev. Mod. Phys. \textbf{86} (2014).
\bibitem{Jav86} J. Javanainen, Oscillatory exchange of atoms between traps containing Bose condensates, Phys. Rev. Lett. \textbf{57}, 3164 (1986).
\bibitem{bra05} T. Brandes, Coherent and collective quantum optical effects in mesoscopic systems, Phys. Rep. \textbf{408}, 315 (2005).


\bibitem{Bre02} H. P. Breuer and F. Petruccione, \textit{The Theory of Open Quantum Systems}, (Oxford University Press, Oxford, 2002).
\bibitem{Car93} H. Carmichael, \textit{An Open Systems Approach to Quantum Optics}, (Springer, Berlin, 1993).

\bibitem{tobe} M. Konopik and E. Lutz, in preparation.
\bibitem{Ama00} S. Amari and H. Nagaoka, \textit{Methods of information geometry} (Oxford University Press, Oxford, 2000), Eq.~(7.61).
\bibitem{Coh04} C. Cohen-Tannoudji, J. Dupont-Roc, and G. Grynberg, \textit{Atom photon interactions}, (Wiley, Weinheim, 2004).
\bibitem{Cam16} S. Campbell, G. De Chiara, and  M. Paternostro, Equilibration and nonclassicality of a double-well potential, Sci. Rep. \textbf{6}, 1 (2016).
\bibitem{sm} See Supplemental Material.
%\bibitem{Shi17} A. Shimizu, K. Fujikura, Quantum violation of fluctuation-dissipation theorem, J. Stat. Mech. 024004 (2017).
%\bibitem{Ma71} W. Marshall, S. W. Lovesey, Theory of thermal neutron scattering, Oxford university press Inc, Oxford, (1971).
%\bibitem{Clo68} Des Cloizeaux et al, Theory of condensed matter, international atomic energy agency, Vienna, (1968).
%\bibitem{Mon13}A. Monras, Phase space formalism for quantum estimation of Gaussian states, arXiv 1303.3682v1 (2013).
\end{thebibliography}
\end{document}